\documentclass[draft]{aipproc}

\layoutstyle{6x9}
\usepackage{amsmath}
\usepackage{amssymb}

\newcommand\GW{gravitational wave}

\begin{document}

\title{GRB-triggered searches for gravitational waves in LIGO data}

\author{Alexander Dietz (LIGO Scientific Collaboration)}{
  address={Cardiff University, Cardiff, CF24 3AA, United Kingdom}
}

\classification{04.30.Tv,04.80.Nn,98.70.Rz}
\keywords      {gravitational-wave astrophysics, gravitational wave detectors and experiments, gamma-ray bursts}

\begin{abstract}
The LIGO \GW{} detectors have recently reached their design sensitivity and finished a two-year science run. 
During this period one year of data with unprecedented sensitivity has been collected. 
I will briefly describe the status of the LIGO detectors and the overall quality of the most recent science run. 
I also will present results of a search for inspiral waveforms in \GW{} data coincident with the short gamma ray burst detected on 1st February 2007, with its sky location error box overlapping a spiral arms of M31. 
No \GW{} signals were detected and a binary merger in M31 can be excluded at the 99\% confidence level.
\end{abstract}

\maketitle

\section{Introduction}

Gamma-ray bursts (GRBs) are intense flashes of gamma-rays which are
observed to be isotropically distributed over the sky~\cite{Klebesadel:1973,Piran:2005,Meszaros:2002} and which fall into two major phenomenological categories, long-duration and
short-duration.
Long duration GRBs have a duration $\gtrsim 2$ seconds and are in general believed to be associated with stellar core-collapse events \cite[e.g.][]{Campana:2006,Malesani:2004,Hjorth:2003jt,Galama:1998ea,Woosley:2006}. 
Short duration GRBs (SGRB) with a duration of $\lesssim$~2 seconds are believed to have a different origin. 
The leading hypothesis is the merger of two compact objects, one of them a neutron star and the other either a neutron star or a black hole \cite[see, for example,][and references therein]{Nakar:2007,Bloom:2007}. Such mergers are also sources of gravitational waves, accessible to ground-based detectors like LIGO, GEO600, and VIRGO
\citep{GRB030329,Acernese:2006,Kochanek:1993,LIGO-Bursts-s3,LIGO-Inspiral-s2-bns,Finn:2004}.
Another possible progenitor of short GRBs are soft gamma repeaters, but at most $\sim$15\% of known short GRBs can be explained by these \citep{Nakar:2006,Chapman:2007}.

\begin{figure}
\includegraphics[height=.35\textheight]{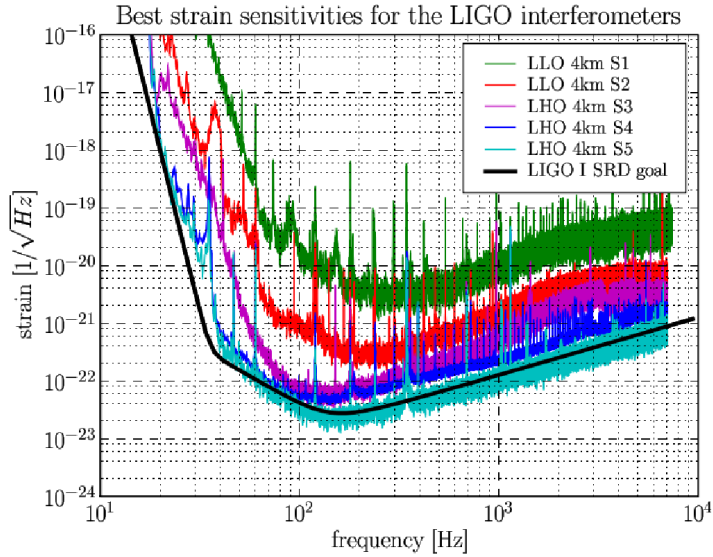}
\caption{\label{f:sensitivity}
Strain sensitivity plot of the most sensitive LIGO detector during the five science runs. 
An improvement of several orders of magnitude (depending on the frequency) can be seen when comparing the first and the latest science run. 
The solid black line represents the design sensitivity, which 
is almost reached at low frequencies and slightly surpassed at higher frequencies.}
\end{figure}

\section{Status of LIGO}
\label{sec:LIGO-Observations}

The Laser Interferometer Gravitational-Wave Observatory (LIGO) consists of three instruments at two different sites, one located near Livingston, LA with a 4~km long detector (referred to as L1) and the other located near Hanford, WA, with a 4~km and 2~km long detector, (referred to as H1 and H2, respectively) \cite{LIGO2007}. 
Also part of the LIGO Scientific Collaboration is GEO600, a British-German instrument located near Hanover, Germany, a 600~m long detector \cite{GEO2007}.

These detectors use suspended mirrors at the ends of kilometer-scale, orthogonal arms to form a power-recycled Michelson interferometer with Fabry-Perot cavities. 
The interference of the light beams from the two arms recombining at the beam splitter depends on the difference between the lengths of the two arms. 
The LIGO detectors are sensitive in a frequency band of $\sim$40~Hz to $\sim$2000~Hz, with the greatest sensitivity at ${\approx}150$~Hz.

\begin{figure}
\includegraphics[height=.3\textheight]{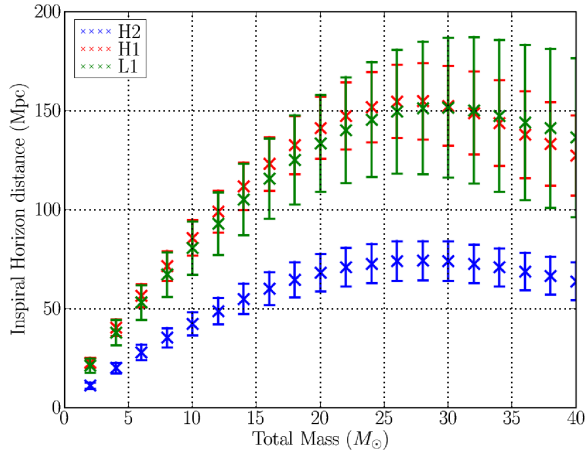}
\caption{\label{f:range}
The horizon distance to an inspiral signal for the three LIGO detectors, as a function of total mass of the merging system. The horizon distance for a neutron star binary merger is about 35~Mpc for each of H1 and L1, our most sensitive detectors.}
\end{figure}

The first data-taking run took place in 2002. The last science run, referred to as S5, started on November 4th, 2005 and ended on October 1st, 2007, collecting more than 1 year of coincident data. 
In May 2007 the French-Italian VIRGO instrument \cite{Acernese:2007}, consisting of a 3~km long detector, joined the S5 data-taking effort. 
Figure \ref{f:sensitivity} shows the strain sensitivity of the most sensitive LIGO detector for each of the five science runs.
In Figure \ref{f:range}, we plot the distance at which an optimally oriented and located source would create a signal-to-noise ratio of 8, the so-called horizon distance, as a function of the total mass of the source.

No \GW{} signals have been detected so far. In particular, results of previous GRB and SGR searches can be found in \citep{GRB030329, SGR1806, BurstS2S3S4}.

\section{Analysis of GRB 070201 for inspiral signals}

GRB~070201 was an intense, short duration GRB, which was detected and localized by five interplanetary Gamma-Ray Burst Timing Network spacecraft (Konus-Wind, Mars Odyssey, INTEGRAL and MESSENGER). The refined error-box of the location of this GRB overlaps with the spiral arms of M31 ~\citep{gcn6098,gcn6103}, making this a possible close-by source of gravitational waves.

At the time of GRB~070201, only the two Hanford detectors were taking data. The horizon distances at this time were 35.7~Mpc and 15.3~Mpc for H1 and H2, respectively.
However, M31 was located sub-optimally at this time so the actual reach is only $\sim 43\%$ of the horizon distance in the direction of M31.
If this GRB were due to a binary binary merger in M31, the \GW{} signal would be easily detectable by LIGO.
A 180~second long on-source segment was chosen around the time of the GRB similar to previous analyses (see e.g. \cite{Finn:2004}), starting 120~seconds before the trigger and lasting 60~seconds after it. 
Off-source segments of the same length were chosen to estimate the rate of accidental background events. By adding simulated signals to two of the off-source segments and analyzing them the efficiency of this search was estimated.

The search for inspiral signals from compact binaries involves matched filtering~\citep{wainstein:1962} the data with theoretical waveforms~\citep{Blanchet:2006av} of two inspiraling objects; the component masses were chosen to cover the region $1~M_\odot < m_1 < 3~M_\odot$ and $1~M_\odot <m_2 < 40~M_\odot$. 
One of the objects must be a neutron star in order to create a GRB ~\citep{Vallisneri:1999nq,Rantsiou:2007ct}, but the other can be either a neutron star or a black hole, which is the reason for the selected mass ranges.
Details of the inspiral search can be found in \cite{LIGO-Inspiral-s2-bns, Anderson:2007,LIGOS2bbh,LIGOS2macho}, and details for this particular search are given in \cite{grb070201}.

\begin{figure}[ht!]
\includegraphics[height=.2\textheight]{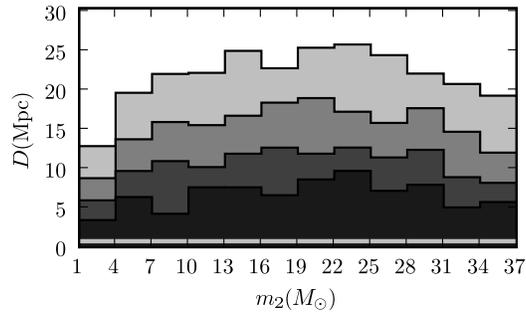}
\caption{\label{f:pcsignal-loudest}
Exclusion areas for an inspiral merger signal from the direction of GRB~070201. The x-axis shows the more massive component of the merger, and the y-axis shows the physical distance. The shaded regions represent $90\%$, $75\%$, $50\%$, and $25\%$ exclusion regions, from darkest to lightest, respectively.
The distance to M31 is indicated by the horizontal gray line at $D=0.77 \textrm{ Mpc}$. }
\end{figure}

No candidates were detected in the on-source time, and therefore no plausible gravitational wave signals were identified around the time of GRB~070201. A mass-dependent exclusion plot on the distance for a binary merger signal is shown in Figure~\ref{f:pcsignal-loudest}. The gray line in this plot indicates the distance of M31 at $0.77\pm0.05$~Mpc. At this distance, a compact binary progenitor is excluded at the $>99\%$ confidence level. 

If, however, the origin of this GRB was a soft-gamma repeater located in M31, the expected \GW{} signal would be too weak to be detected by LIGO, so we cannot exclude it \cite{Mazets:2007}.
A model independent analysis looking for a burst of gravitational waves was also performed on the same data, but it could not exclude a possible SGR at the distance of $D_\textrm{M31}\simeq 770$~kpc \cite{grb070201}.
Therefore, our non-observation of a \GW{} signal associated with GRB~070201 suggests that the progenitor could be an SGR in M31 or a binary merger at a much greater distance.

\section{Discussion}

The LIGO detectors have recently finished a science run at design sensitivity, collecting one year of coincident data. During this period, several missions detected GRB 070201, a short-duration GRB, and provided a localization region that overlapped M31.
Data from the LIGO H1 and H2 gravitational-wave detectors were analyzed around the time of this GRB, looking for inspiral waveforms over a range of component masses. No plausible gravitational-wave signals were identified.
Based on this search, a compact binary progenitor (neutron star + black hole or a binary neutron star) located in M31  
could be excluded at the 99\% confidence level. 
A burst search looking for an un-modeled burst of gravitational waves on the same data was performed, with the conclusion that less than 7.9$\times 10^{50}$~ergs was emitted within any 100-ms-long time interval inside the on-source region at the most sensitive frequency of LIGO ($f\approx$~150~Hz), if the source were located in M31. 

The search for inspiral \GW{} signals, associated with short GRBs will continue on all (short) GRBs detected during S5. 
Since only 7 of them have a redshift estimation, the real distance to the GRBs is not known and could be within the reach of the LIGO/VIRGO detectors. Results from this search will be reported in the following months.

\section{Acknowledgement}
The authors gratefully acknowledge the support of the United States
National Science Foundation for the construction and operation of the
LIGO Laboratory and the Science and Technology Facilities Council of the
United Kingdom, the Max-Planck-Society, and the State of
Niedersachsen/Germany for support of the construction and operation of
the GEO600 detector. The authors also gratefully acknowledge the support
of the research by these agencies and by the Australian Research Council,
the Council of Scientific and Industrial Research of India, the Istituto
Nazionale di Fisica Nucleare of Italy, the Spanish Ministerio de
Educaci\'on y Ciencia, the Conselleria d'Economia, Hisenda i Innovaci\'o of
the Govern de les Illes Balears, the Scottish Funding Council, the
Scottish Universities Physics Alliance, The National Aeronautics and
Space Administration, the Carnegie Trust, the Leverhulme Trust, the David
and Lucile Packard Foundation, the Research Corporation, and the Alfred
P. Sloan Foundation.
This paper was assigned LIGO Document Number: LIGO-P080002-00-Z.


\bibliographystyle{aipproc}   

\bibliography{/home/alex/Work/Publications/Papers/ProcGRB/ProcGRB.bib}

\end{document}